\documentclass[twocolumn]{aastex62}

\newcommand{\arcs}{$^{\prime\prime}$} 
\newcommand{\beq}{\begin{equation}\begin{aligned}}
\newcommand{\eeq}{\end{aligned}\end{equation}}

\graphicspath{{./}{figures/}}

\accepted{\today}
\submitjournal{ApJL}

\shorttitle{Satellite System of M101}
\shortauthors{Carlsten et al.}

\begin{document}

\title{Using Surface Brightness Fluctuations to Study Nearby Satellite Galaxy Systems: the Complete Satellite System of M101}

\correspondingauthor{Scott G. Carlsten}
\email{scottgc@princeton.edu}

\author[0000-0002-5382-2898]{Scott G. Carlsten}
\affil{Department of Astrophysical Sciences, 4 Ivy Lane, Princeton University, Princeton, NJ 08544}

\author[0000-0002-1691-8217]{Rachael L. Beaton}
\altaffiliation{Hubble Fellow}
\affiliation{Department of Astrophysical Sciences, 4 Ivy Lane, Princeton University, Princeton, NJ 08544}
\affiliation{The Observatories of the Carnegie Institution for Science, 813 Santa Barbara St., Pasadena, CA~91101\\}

\author[0000-0003-4970-2874]{Johnny P. Greco}
\altaffiliation{NSF Astronomy \& Astrophysics Postdoctoral Fellow}
\affiliation{Center for Cosmology and AstroParticle Physics (CCAPP), The Ohio State University, Columbus, OH 43210, USA}

\author{Jenny E. Greene}
\affil{Department of Astrophysical Sciences, 4 Ivy Lane, Princeton University, Princeton, NJ 08544}

\begin{abstract}
We use surface brightness fluctuation (SBF) measurements to constrain the distance to low surface brightness (LSB) dwarfs in the vicinity of M101. Recent work has discovered many LSB candidate satellite companions of M101. However, without accurate distances, it is problematic to identify these dwarfs as physical satellites of M101. We use CFHT Legacy Survey (CFHTLS) data to measure the SBF signal for 43 candidate dwarfs. The data is deep enough that we constrain 29 of these to be unassociated background galaxies by their lack of SBF. We measure high S/N SBF signals for two of the candidate dwarfs, which are consistent with being at the distance of M101. The remaining candidates are too LSB and/or small for their distances to be constrained. Still, by comparison with Local Group dwarfs, we argue that the M101 satellite system is likely now complete down to stellar masses of $\sim5\times10^5$ M$_\odot$. We also provide a new SBF distance for the nearby dwarf UGC 8882, which suggests that it is significantly outside of the virial radius of M101 and is thus not a physical satellite. By constraining the distances to a majority of the candidates using only archival data, our work demonstrates the usefulness of SBF for nearby LSB galaxies and for studying the satellite systems of nearby massive galaxies.
\end{abstract}


\section{Introduction}
Expanding the census of faint, nearby dwarf galaxies is crucial to understand structure formation on the smallest scales. In recent years this has largely been done in the form of characterizing the dwarf satellite systems of nearby ($D<20$ Mpc) Milky Way (MW) analogs \citep[e.g.][]{muller101, smercina2018, danieli101, geha2017, bennet2017, park2017, kim2011, merritt2014} or LMC analogs \citep[e.g.][]{madcash} with the goal of addressing the small-scale problems in $\Lambda$CDM \citep[e.g.][]{bullock2017}. Generally, these studies find and catalog low surface brightness (LSB) objects in deep, wide-field imaging and then either determine the distance to these objects (perhaps with follow-up spectroscopy or HST observations) to confirm association with a host or group or simply assume association based on proximity on the sky. This latter assumption is very problematic for nearby systems that might be contaminated by a background group in the same area of the sky \citep[e.g.][]{merritt2016, danieli101, cohen2018}.

In a companion paper (Carlsten et al. submitted) we show that ground-based surface-brightness fluctuation (SBF) measurements can efficiently provide distances and, hence, confirm association for many LSB dwarfs using the same images in which the objects were discovered. In addition, we provide an absolute SBF calibration and show that distances of 15\% accuracy are possible for dwarfs as low SB as $\mu_{0i}\sim25$ mag arcsec$^{-2}$ in $\sim$ 1 hour exposure time with CFHT. Our calibration is credible over the range $0.3\lesssim g-i\lesssim0.8$ mag.

In this Letter, we catalog, using SBF,
the dwarf companions of M101 (NGC 5457). M101 is a nearby (D=7Mpc; \citet{m101_dist, t15}), massive spiral galaxy with peak circular velocity of $\sim210$ km/s \citep{sofue_m101_vcirc} which makes it a close analog in mass to the MW. It exhibits a minor pseudobulge, which contributes 3\% of the luminosity \citep{kormendy2010}, indicating a relatively merger-free history. This is corroborated by its anomalously faint stellar halo \citep{pvd_m101_halo}. These features make its satellite system an interesting target of study to address predictions from structure formation models on the correlation between bulge mass and satellite abundance \citep[e.g.][]{lopez_bulge, henkel, javanmardi_bulge}.

Early studies of M101's satellite system indicated a low abundance of satellites and almost no dwarf ellipticals or dwarf spheroidals \citep{bremnes}. In more recent work, the M101 satellite system has been surveyed by different groups, including using SDSS data, CFHTLS data, and two different small telescope surveys. Many LSB objects have been found and catalogued but very few have any distance constraints. In this paper, we measure the SBF signal for many of these candidate satellites to constrain the distance, either showing them to be background or actual satellites.

This Letter is organized as follows: in \S\ref{sec:data} we describe the galaxy sample and data used, in \S\ref{sec:sbf} we present the SBF distance measurements, in \S\ref{sec:discussion} we discuss the results, and summarize in \S\ref{sec:sum}.

\section{Galaxy Sample and Data}\label{sec:data}
We primarily use the catalog of candidate companions of \citet{bennet2017} who used the CFHTLS. Due to the depth and resolution of the CFHTLS data, this catalog superseded the previous catalogs as it recovers all the previously discovered objects. Additionally, the detection algorithm of \citeauthor{bennet2017} is automated with well-understood incompleteness. This catalog includes the objects discovered by \citet{merritt2014} and \citet{k15}\footnote{\citet{javanmardi_m101} independently discovered one of the objects (Dw A) of \citet{k15} using the same small-telescope dataset.}, in addition to several new discoveries. For completeness, we include the seven Dragonfly objects from \citet{merritt2014} in this analysis as well even though they have distance constraints from HST TRGB. \citet{danieli101} presented HST TRGB distances for three of these (M101-DF1, M101-DF2, and M101-DF3), demonstrating they are at the distance of M101, and \citet{merritt2016} showed the remaining four were at least twice the distance of M101 and likely associated with the massive elliptical NGC 5485 at a distance of 27 Mpc. However, only one of the other objects in the catalog of \citet{bennet2017} has any distance information\footnote{Dw26 of \citet{bennet2017} is known to be D $\sim$ 150Mpc from H\,{\sc i} observations.}. Two of the objects of \citet{muller101}, dw1408+56 and dw1412+56, are in the CFHTLS footprint and we include those objects as well.

Additionally, we measure the SBF signal for the bright dwarf UGC 8882. UGC 8882 has a previous SBF distance from \citet{jerjen_field2} using the calibration of \citet{jerjen_field}. We provide an updated SBF distance based on the much more robust empirical calibration of Carlsten et al. (submitted).

\begin{figure*}
\includegraphics[width=\textwidth]{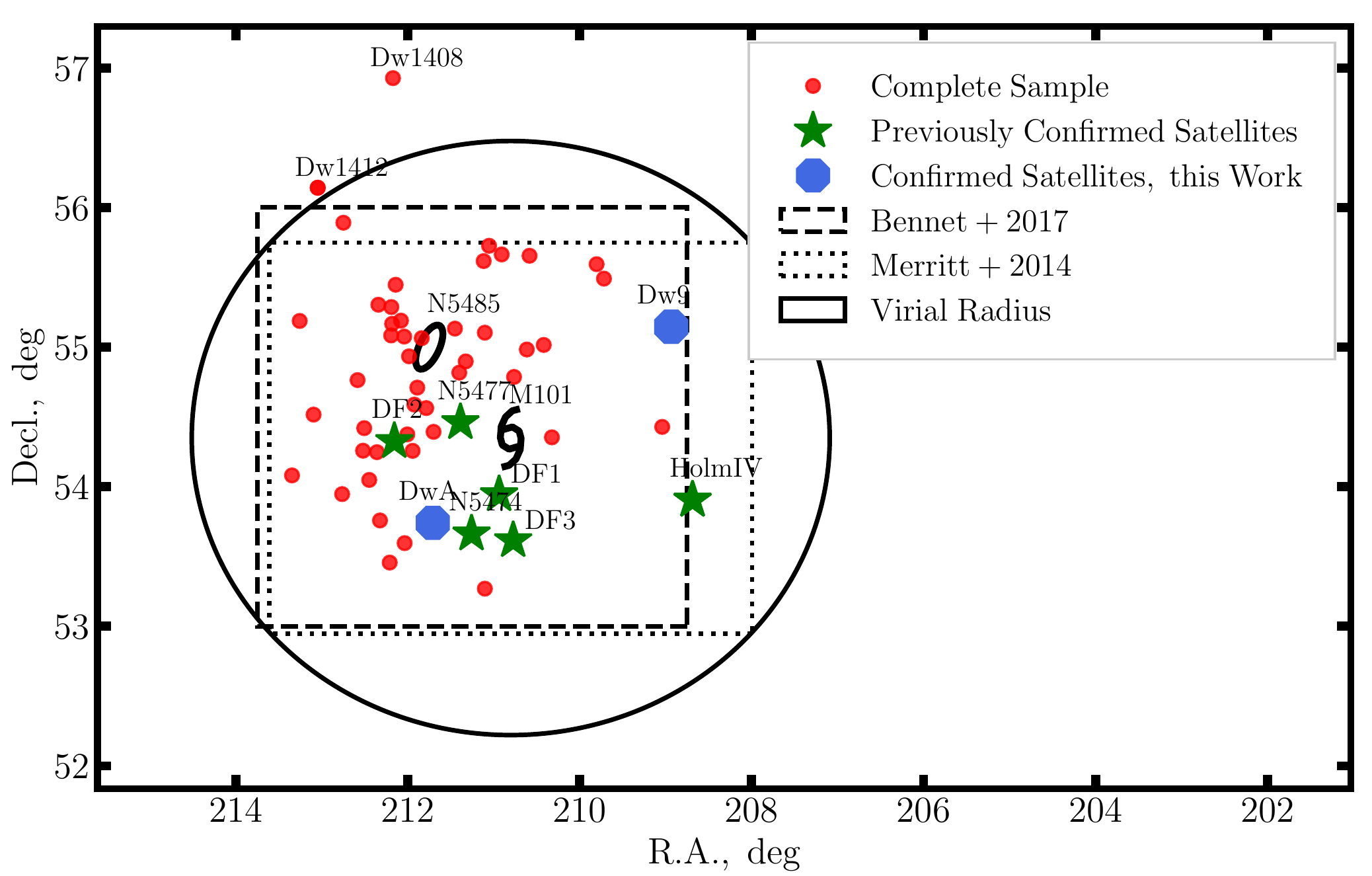}
\caption{The sample of LSB dwarfs analyzed in this work shown relative to M101 and the background massive elliptical NGC 5485 at D=27 Mpc \citep{merritt2014}. The virial radius of M101 of 260 kpc is also shown relative to the footprints of two different surveys of the region. The figure shows the previously confirmed satellites along with two satellites confirmed in the current work.}
\label{fig:m101_sys}
\end{figure*}

We point the reader to Table 1 of \citet{bennet2017} for locations and properties of the candidate dwarfs.  Figure \ref{fig:m101_sys} shows the layout of the sample relative to the footprints of the different surveys used and M101's virial radius. The candidate dwarfs appear to cluster around NGC 5485, suggesting many are unassociated with M101. 

In this project, we use MegaCam \citep{megacam} data from the CFHT taken as part of the Legacy Survey Wide layer. The CFHTLS-Wide data have a characteristic 50\% completeness depth of 26-26.5 mag in $g$ and 25.5-26.0 mag in $i$ \citep{cfhtls}. As discussed in Carlsten et al. (submitted), the default MegaCam pipeline \citep{megapipe} sky subtraction is unsuitable for measuring SBF of LSB objects. Instead, we download the \textsc{Elixir} \citep{elixir} pre-processed CCD-level frames from the CADC archive\footnote{\url{http://www.cadc-ccda.hia-iha.nrc-cnrc.gc.ca/en/}} and perform our own photometric and astrometric calibration, sky-subtraction, and stacking. We acquire and reduce images in both $g$ and $i$ band.

\section{SBF Distances}\label{sec:sbf}
\subsection{SBF Measurement}
We follow the SBF measurement methodology described in detail in Carlsten et al. (submitted) which largely follows the standard SBF process \citep[e.g.][]{tonry2001, blakeslee2009, cantiello2018}. SBF is measured in the $i$ band and the galaxy's $g-i$ color is used to account for the dependence of SBF on the stellar population. In brief, we first fit each galaxy with a S\'{e}rsic profile to model the underlying light profile. \texttt{Imfit} \citep{imfit} is used to do the fitting. Carlsten et al. (submitted) performed image simulations of LSB galaxies and found that the sky subtraction algorithm used here allowed the colors of the galaxies to be recovered with roughly 0.1 mag accuracy. We take this as a characteristic uncertainty in the color measurements. These smooth profile models are then subtracted from the image and we mask nearby foreground stars and background galaxies. We use \texttt{sep}\footnote{\url{https://github.com/kbarbary/sep}} \citep{sep}, a Python implementation for SExtractor \citep{sextractor}, for the object detection. Thresholds for detection were generally in the range 2$\sigma$ to 5$\sigma$ above the background. The threshold was adjusted on a per galaxy basis to ensure that clear contaminating sources were always masked. These thresholds correspond to absolute magnitudes of $M_i\lesssim-4$ at the distance of M101. This ensures that star clusters associated with the galaxy are masked but the RGB tip stars in the dwarf galaxies are not. 

Once masked, the images are normalized by the square root of the smooth galaxy model image and a Fourier transform is taken to calculate the power spectra. The actual region included in the Fourier transform is an ellipse centered on the galaxy that extends out to the radius where the galaxy profile drops below $\sim0.3$ times its maximum level. The azimuthally averaged power spectrum is fit with a combination of the PSF power spectrum convolved with the mask power spectrum and a constant, representing the contribution of white noise to the image fluctuations.

\begin{figure*}
\includegraphics[width=0.97\textwidth]{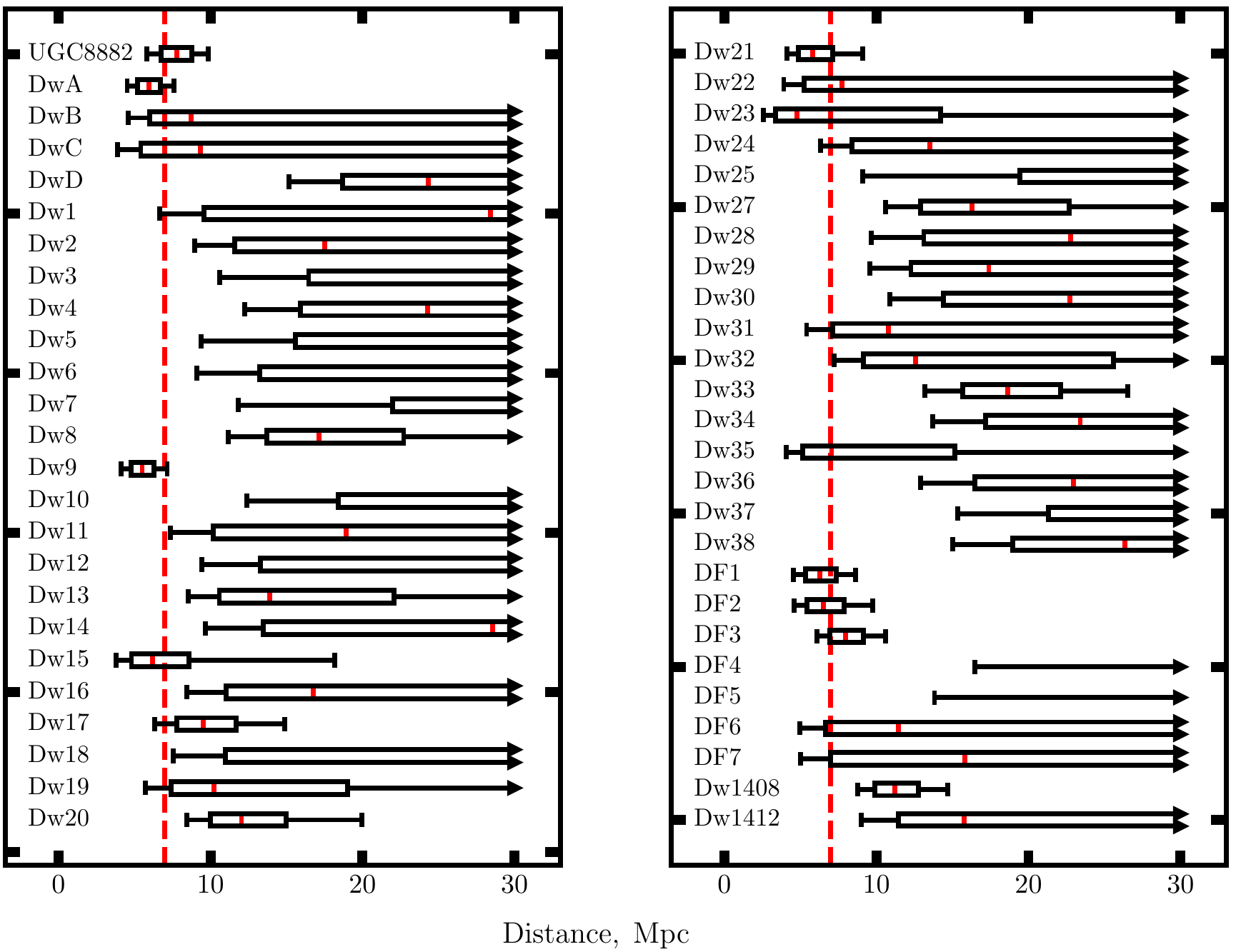}
\caption{Box and whisker plot showing percentiles of the distance distribution that we derive for each galaxy in the sample. The whiskers extend from the 2.5$^{th}$ percentile to the 97.5$^{th}$ ($\pm2\sigma$) and the boxes extend from the 16$^{th}$ percentile to the 84$^{th}$ ($\pm1\sigma$). The red mark inside the boxes denotes the median. Arrows pointing to the right indicate that the distribution extends to larger distances. The dashed vertical red line at 7 Mpc indicates the distance of M101.}
\label{fig:dist_bounds}
\end{figure*}

As described in Carlsten et al. (submitted), the uncertainty in the SBF measurement comes from two major sources. We estimate the uncertainty coming from the actual power spectrum fit by varying the range of wavenumbers used in the fit and the region of the galaxy used in a Monte Carlo approach. From this, we get a median fluctuation level and an uncertainty from the spread in measured fluctuations. The second main source of uncertainty comes from contamination from residual, unmasked sources. For this, we measure the SBF signal in nearby background fields around each galaxy. These fields have undergone the same normalization and masking as the galaxy. We determine a median residual fluctuation level and uncertainty from the spread in the ensemble of fields used.

\subsection{Bounds on Distance}
With measured SBF signals in hand, we turn to extracting distance information for the dwarfs in our sample. Our goal is not necessarily to determine distances for each dwarf because, as shown below, the SBF signal is very weak (or nonexistent) for most of the dwarfs, making an SBF distance impossible. Instead, the goal is to set lower bounds on the distance based on the SBF signal or lack thereof.

To determine bounds on the distance from the measured SBF signal, we use the empirical absolute calibration of Carlsten et al. (submitted). We start with the measured fluctuation signal from the $i$ band image for each galaxy and propagate uncertainties in the SBF measurement and SBF calibration in a Monte Carlo approach. For each of 10,000 iterations, we resample the SBF signal using the measured SBF signal and its uncertainty (assuming Gaussian distributions) and resample the residual SBF signal using its measured value and spread. We subtract the residual signal from the signal measured from the galaxy and calculate the apparent SBF magnitude using the photometric zeropoint of the images. The color of the galaxy is similarly resampled using the measured value and spread and used in the SBF calibration of Carlsten et al. (submitted) to calculate the absolute SBF magnitude. From this a distance modulus and distance are calculated. We then calculate the 2.5$^{th}$, 16$^{th}$, 50$^{th}$, 84$^{th}$, and 97.5$^{th}$ percentiles from the distribution of distances for each galaxy. For many of the galaxies, the resampled SBF signal could be zero or less than zero. For these galaxies, the distance distribution extends to infinity but a lower bound on the distance is still possible. Figure \ref{fig:dist_bounds} shows these distance percentiles for each galaxy in our sample. %

\begin{figure*}
\includegraphics[width=\textwidth]{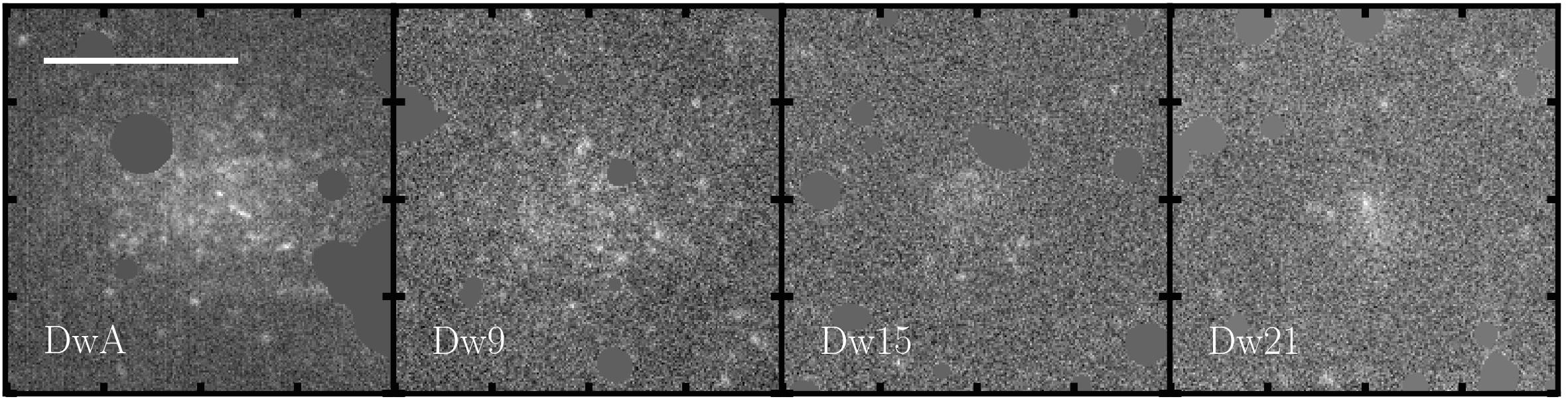}
\caption{The four candidate dwarfs that have tight ($\pm1$ Mpc) distance constraints that put them at the distance of M101. The $i$ band images are shown, masked by the mask used in the SBF analysis. The white bar in the upper left corner is 10\arcs (each image is at the same angular scale).}
\label{fig:candidates}
\end{figure*}

We note four rough groups of galaxies in Figure \ref{fig:dist_bounds}. First are the galaxies that have very wide distributions in distance but have 2$\sigma$ lower bounds in distance beyond M101. We can conclude that these galaxies are background because they would have measurable SBF signal if they were at the distance of M101, which is not observed. Another group are the galaxies which have distance distributions extending from within M101's distance out to very large distances. These are generally the faintest and smallest galaxies of the sample and very little can be said about their distances since the SBF would be too faint to observe even if the galaxies were at the distance of M101. A third group are the galaxies that seem to have narrow distance distributions at distances beyond M101 (e.g., Dw1408, Dw20, and Dw33). These galaxies exhibit non-S\'{e}rsic shapes and the residuals from using a S\'{e}rsic profile as a model for the smooth background could be adding spurious fluctuations into the SBF measurement. In these cases, we do not fully trust the SBF distances. The conclusion that they are beyond M101 is robust, however, because even with the added fluctuation power from an improper fit, they do not exhibit as much brightness fluctuation as they should if they were at the distance of M101. The final group of galaxies are those that have narrow distance distributions (with $\pm1\sigma$ distance ranges of $\lesssim2$ Mpc) centered on the distance of M101. These are particularly exciting as they are possibly satellites of M101 and include DwA, Dw9, Dw15, and Dw21. We discuss these objects in more detail in the next section.

We recover the distances of M101-DF1, M101-DF2, and M101-DF3 to be at the distance of M101. This agrees with the HST TRGB analysis of \citet{danieli101} and gives confidence in our SBF measurements\footnote{We note, however, that M101-DF1, M101-DF2, and M101-DF3 were among the galaxies used in Carlsten et al. (submitted) to derive the calibration used here.}. We are able to show that DF4 and DF5 are background but could not say anything about DF6 and DF7 due to their extreme faintness. Our distance of 7.8$\pm1.0$ Mpc for UGC 8882 is consistent with the distance of 8.3$\pm0.8$ Mpc that \citet{jerjen_field2} report.

\subsection{Confirmed Satellites}\label{sec:confirmed}
As mentioned above, four of the dwarfs appear to have significant SBF signals that put them at the distance of M101. Figure \ref{fig:candidates} shows the $i$ band images of these four candidates. We note that DwA and Dw9 appear semi-resolved into stars. The SBF is very strong in both galaxies (S/N of 15 and 9, respectively). The other two have weaker SBF signals (S/N $\sim$ 2-3). It is possible that the high observed fluctuation power is coming from residuals in the S\'{e}rsic fitting or, in the case of Dw21, from a single unmasked bright source in the galaxy. We take the conservative approach and include these two galaxies in the group of galaxies that do not have firm distance constraints but note they could be high priority targets for deeper follow-up. In total, out of the 43 galaxies in our sample that had no previous distance information\footnote{Therefore not including any Dragonfly objects, Dw26, or UGC 8882.}, we demonstrate that 29 are background and 2 are likely satellites of M101. For the remaining 12, we are unable to constrain the distance from the current data.

\begin{figure}
\includegraphics[width=0.47\textwidth]{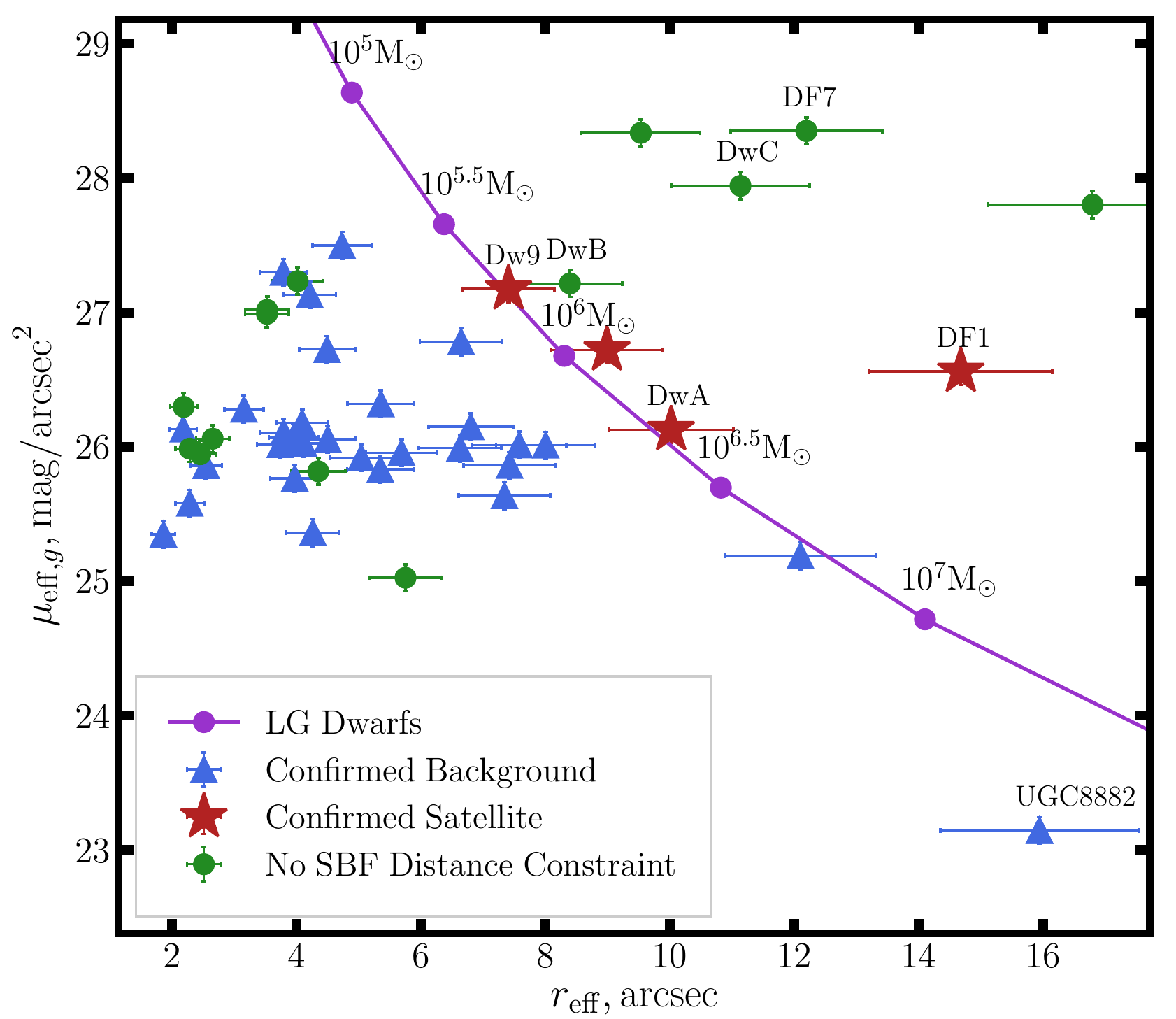}
\caption{The surface brightness at the effective radius and effective radii of the galaxy sample, sorted by whether the SBF measurement alone could constrain them to be background or actual satellites or whether no constraint was possible. The purple curve shows the surface brightness and size for different stellar mass dwarfs from the mean relations for LG dwarfs given in \citet{danieli_field} (at the distance of M101). The mean surface brightness within the effective radius from \citet{danieli_field} is converted to a surface brightness at the effective radius assuming a n=0.7 S\'{e}rsic profile. A 10\% error in the effective radius is assumed for each point. }
\label{fig:gal_prop}
\end{figure}

\section{Discussion}\label{sec:discussion}

\subsection{Completeness of the Satellite System}
To explore the properties of the galaxies that we can set distance constraints versus those that we cannot, we plot the surface brightness and effective radii of the sample in Figure \ref{fig:gal_prop}. The galaxies are split into three groups: those confirmed to be background with the SBF, those confirmed to be satellites with the SBF, and those where a distance constraint was not possible. The large, bright galaxies are generally those with a distance constraint, as expected. We also show the expected size and surface brightness as a function of dwarf stellar mass from the mean relations for Local Group dwarfs from \citet{danieli_field}. 
Most of the galaxy sample is smaller than the LG dwarfs at the same surface brightness because, as we are finding, most are background.

The curve for the LG dwarfs appears to leave the region where distance constraints are possible at a stellar mass of $\sim5\times10^5$ M$_\odot$. This appears to be roughly the stellar mass of the dwarf Dw9. From Figure 2 of \citet{bennet2017}, the catalog of candidate satellites should be complete at $r_{\mathrm{eff}}\sim8$ \arcs~ down to central surface brightness of $\mu_{0,g}\lesssim26$ mag/arcsec$^2$. This corresponds to a surface brightness at the effective radius of $\mu_{\mathrm{eff},g}\lesssim28$ mag/arcsec$^2$ for the $n=1$ S\'{e}rsics used by \citet{bennet2017}. This is about a magnitude fainter than the effective limit of SBF for getting a distance constraint, so the catalog of dwarfs should be complete for dwarfs of similar mass to Dw9. From this and the fact that the CFHTLS data covers most of the virial volume of M101 (cf. Figure \ref{fig:m101_sys}), we argue that the satellite system of M101 is likely now complete down to stellar masses of $\sim5\times10^5$ M$_\odot$.

\begin{deluxetable}{ccccc}
\tablecaption{Confirmed satellites of M101}

\tablehead{\colhead{Name} & \colhead{R.A.} & \colhead{Decl.} & \colhead{Distance} & \colhead{} \\ 
\colhead{} & \colhead{} & \colhead{} & \colhead{(Mpc)} & \colhead{} } 
\startdata
NGC 5474 & 14:05:01.6 & +53:39:44 & 6.82$\pm0.41$$^a$ &  \\
NGC 5477 & 14:05:33.3 & +54:27:40 & 6.77$\pm0.40$$^a$ &  \\
UGC 8837 & 13:54:45.7 & +53:54:03 & 6.93$\pm0.48$$^a$ &  \\
UGC 9405 & 14:35:24.1 & +57:15:21 & 6.30$\pm0.38$$^a$ &  \\
M101-DF1 & 14:03:45.0 & +53:56:40 & 6.37$^{0.35}_{0.35}$$^b$ &  \\
M101-DF2 & 14:08:37.5 & +54:19:31 & 6.87$^{0.21}_{0.30}$$^b$ &  \\
M101-DF3 & 14:03:05.7 & +53:36:56 & 6.52$^{0.25}_{0.27}$$^b$ &  \\
M101-DwA & 14:06:50.0 & +53:44:29 & 6.0$\pm0.7$$^c$ &  \\
M101-Dw9 & 13:55:44.6 & +55:08:45 & 5.5$\pm0.8$$^c$ &  \\
\enddata
\label{tab:sats}

\tablerefs{$^a$ \citet{t15}, $^b$ \citet{danieli101}, $^c$ Current Work}
\end{deluxetable}

\subsection{Known Satellites}
In Table \ref{tab:sats} we list the known, confirmed satellites of M101, including the two confirmed by the current work. Our sample derives from \citet{t15} and \citet{danieli101}. \citet{k94}, \citet{bremnes}, and \citet{muller101} considered many more nearby galaxies as associates of M101. However, \citet{t15} argued that many of these members (e.g. NGC 5585 and UGC 8882) were, in fact, background/foreground and physically unrelated to M101. We have found a distance of 7.8$\pm1.0$ Mpc for UGC 8882 which is consistent with the D=7 Mpc we have used for M101. However, most of the uncertainty in the distance for UGC 8882 comes from the 0.1 mag uncertainty in color that we assume. Since UGC 8882 is so bright, this is probably overly conservative and a $\pm0.5$ Mpc uncertainty in the distance is more realistic. Recent TRGB work (Beaton et al., in prep) suggests a closer distance to M101 of $\sim$6.4~Mpc, which indicates that UGC 8882 is likely background and not directly a satellite of M101. We note that all the confirmed satellites are closer than 7 Mpc with a median distance of 6.5 Mpc, which supports a closer distance for M101 than either the 7.24 Mpc of \citet{m101_dist} or 6.79 Mpc of \citet{t15}.

We include the galaxy UGC 9405 (DDO 194) in this list, but at 600 kpc from M101, it is outside of the virial radius of M101 \citep[$\sim$260 Mpc][]{merritt2014}. We show the currently known system in Figure \ref{fig:m101_sys}. UGC 9405 is far off the plot to the upper left and its inclusion in the group is questionable. The other satellites exhibit an interesting asymmetry with a majority of the satellites being on one side of M101. A similar asymmetry is seen in M31 \citep{mcconnachie06, conn13} with a significant majority of M31's satellites being on the near side of M31 to the MW. We leave further exploration of this feature to further work. 

We note there are eight known satellites of M101 with more stellar mass than $\sim5\times10^5$ M$_\odot$ (not including UGC 9405). The compilation of MW satellites of \citet{lg_dwarfs} includes eight MW satellites in this mass range (Canis Major, Sagittarius, LMC, SMC, Sculptor, Fornax, Leo\,{\sc i}, Leo\,{\sc ii}). We leave further comparison with the MW satellites to future work.

\section{Summary}\label{sec:sum}
In this contribution, we have demonstrated the effectiveness of using SBF measurements to constrain the distance to LSB dwarf galaxies. We have taken existing catalogs of possible dwarf satellites of the nearby spiral M101 and measured the SBF signal on the same data used in the discovery. For 29 out of the 43 dwarfs in the sample that had no previous distance constraint, we have shown that the galaxies must be background due to their lack of measurable SBF. For two galaxies in the sample, we measured SBF with high S/N which placed them at the distance of M101. The remaining galaxies in the sample were either too faint or too small for the SBF measurement to say anything firm about the distance. 

Since we utilized the same dataset used in discovering the dwarfs, we avoided the need for follow-up to determine distances. If TRGB were used, HST follow-up would likely be required, which for 43 candidates with limited multiplexing and 1-2 orbits per object would be very expensive. At the same time, the fact that at least 29 out of the 43 dwarf candidates are background objects highlights the need for distance measurements when studying the satellite systems of nearby galaxies. 

By comparison with the size and surface brightness of LG dwarfs, we argued that SBF distance constraints were possible with these data down to stellar masses of $\sim5\times10^5$ M$_\odot$. \citet{bennet2017} showed that the candidate catalog is complete at this size and surface brightness. Therefore, since the CFHTLS data used covers most of the virial volume of M101, we argued that the satellite system of M101 is likely complete down to $\sim5\times10^5$ M$_\odot$. Table \ref{tab:sats} lists the known members. This completeness will make M101 useful in confronting predictions from structure formation theories on expected satellite abundance and properties.

Finally, we mention that this sort of analysis will be very useful in the future with large surveys like the Hyper Suprime-Cam\footnote{\url{https://hsc.mtk.nao.ac.jp/ssp/}} \citep{hsc} and LSST. The combined depth and wide area of these surveys will facilitate the discovery of many LSB objects \citep[e.g.][]{johnny1}. Follow-up with HST or JWST for everything discovered will not be possible, but the depth and quality of the survey imaging will make the SBF approach, like we used here, very feasible.

\section*{Acknowledgements}
Support for this work was provided by NASA through Hubble Fellowship grant \#51386.01 awarded to R.L.B.by the Space Telescope Science Institute, which is operated by the Association of  Universities for Research in Astronomy, Inc., for NASA, under contract NAS 5-26555. J.P.G. is supported by an NSF Astronomy and Astrophysics Postdoctoral Fellowship under award AST-1801921. J.E.G. and S.G.C. are partially supported by the National Science Foundation grant AST-1713828.

Based on observations obtained with MegaPrime/MegaCam, a joint project of CFHT and CEA/IRFU, at the Canada-France-Hawaii Telescope (CFHT) which is operated by the National Research Council (NRC) of Canada, the Institut National des Science de l'Univers of the Centre National de la Recherche Scientifique (CNRS) of France, and the University of Hawaii. This work is based in part on data products produced at Terapix available at the Canadian Astronomy Data Centre as part of the Canada-France-Hawaii Telescope Legacy Survey, a collaborative project of NRC and CNRS.

\bibliographystyle{aasjournal}
\bibliography{calib}

\end{document}